\documentclass{article}
\usepackage{spconf,amsmath,graphicx}
\usepackage{multirow}
\usepackage{booktabs}
\usepackage[textfont=it,tableposition=top]{caption}
\usepackage{color}

\title{Large-scale ASR Domain Adaptation Using Self- and Semi-supervised Learning}
\name{
\parbox{.95\linewidth}{\centering
Dongseong Hwang, Ananya Misra, Zhouyuan Huo, Nikhil Siddhartha, Shefali Garg, David Qiu, Khe Chai Sim, Trevor Strohman, Françoise Beaufays, Yanzhang He}}
\address{Google LLC, USA}
\begin{document}
%
\maketitle
\begin{abstract}
Self- and semi-supervised learning methods have been actively investigated to reduce labeled training data or enhance model performance. However, these approaches mostly focus on in-domain performance for public datasets. In this study, we utilize the combination of self- and semi-supervised learning methods to solve unseen domain adaptation problems in a large-scale production setting for online ASR model. This approach demonstrates that using the source domain data with a small fraction of the target domain data (3\%) can recover the performance gap compared to a full data baseline: 13.5\% relative WER improvement for target domain data.
\end{abstract}
\section{Introduction}
\label{sec:intro}

It is natural to have poor speech recognition accuracy with small labeled dataset. There are 2 major approaches to resolving this problem: self-supervised learning (self-sup) and semi-supervised learning (semi-sup). Self-sup pre-trains the audio encoder to learn meaningful representations. Contrastive Predictive Coding (CPC)~\cite{oord2018representation}, Wav2vec~\cite{Schneider2019_wav2vec} and Wav2vec2.0~\cite{baevski2020wav2vec} show that contrastive loss can be good pre-training objectives. Autoregressive Predictive Coding (APC)~\cite{Chung2020_apc} shows that mean squared error (MSE) also can be a good objective. Semi-sup trains the ASR model using the pseudo labels generated by a teacher model. Noisy student training (NST)~\cite{xie2020self} is a popular semi-supervised learning approach, which also works for ASR~\cite{Park_2020, xiao2021contrastive}. Recently, it is also shown that combining self- and semi-sup methods improves the performance of automatic speech recognition (ASR) \cite{zhang2020pushing} significantly, leading to the state-of-art (SOTA) Librispeech Word Error Rate (WER). However, the current methods focused on maximizing the performance for in-domain data. The largest dataset of the studies is labeled Librispeech data (1k hours) \cite{panayotov2015librispeech} and unlabeled LibriLight data (60k hours) \cite{kahn2020libri}.

In large-scale production ASR systems, there is often the challenge of mismatched domain between the training data (source domain) and the real world data (target domain) \cite{9296327}. It is also common to have an order of magnitude bigger dataset with various contexts (e.g. command, chat, caption) than the public dataset like Librispeech. To the best of our knowledge, there are no extensive studies to tackle large-scale domain adaptation using both self- and semi-supervised learning.

In this paper, we propose a combined self- and semi-sup approach for domain adaptation. Our method completely recovers target domain accuracy using only a small fraction of labeled target data. In addition, the improved generalization power enhances source domain accuracy as well. The other main contributions of our work is that we analyze how much self- and semi-sup contribute on domain adaptation. Self-supervised learning improves overall model generalization and semi-sup directly closes out-of-domain generalization gap, which means both are complementary.

\section{Related works}
\label{sec:rel_work}

We use the three popular self-supervised learning methods in this work: Wav2vec \cite{oord2018representation, Schneider2019_wav2vec}, and  Wav2vec2.0 \cite{baevski2020wav2vec}, APC \cite{Chung2020_apc}.

The Wav2vec loss maximizes the mutual information between the context latent vector and the future inputs \cite{Schneider2019_wav2vec}. Instead of the convolution layers, we use conformer blocks~\cite{DBLP:conf/interspeech/GulatiQCPZYHWZW20} for the context network. As we directly use the log-mel features instead of the waveform, we exclude the feature network. The output of the audio encoder directly predicts the future log-mel features.

The Wav2vec2.0 loss maximizes the mutual information between the context latent vector and the masked input features \cite{baevski2020wav2vec}. We exclude the feature network for the same reason. We do not quantize the input features as we do not observe any performance difference in our experiments.

The APC loss minimizes the MSE between the input log-mel feature and the predicted log-mel feature by the audio encoder \cite{Chung2020_apc}. APC paper uses LSTM or transformer blocks but we use conformer blocks.

Noisy student training (NST) \cite{Park_2020} is one of the most popular semi-sup methods. It trains a model with both labeled data and unlabeled data. The teacher model produces pseudo labels of unlabeled data from non-augmented input feature. The method trains the student model with augmented input feature and the pseudo labels as the ground truth. The current Librispeech SOTA paper \cite{zhang2020pushing} uses NST with Librispeech data (labeled data) and LibriLight data (unlabeled data). Unlike image domain, it's hard to use soft label for sequence model. NST in speech domain \cite{Park_2020, zhang2020pushing} uses hard label.

\section{Methods}
\label{sec:methods}

\subsection{Self-Supervised learning}
\label{sec:selfsup}
All of the self-supervised methods are used to pre-train the audio encoder of the RNN-T model \cite{Graves2012} using all the source and target domain data. This is followed by supervised training of the entire model using only the labeled source domain data. 
However, with this 2-stage training approach, it is difficult to determine the optimal pre-training hyper parameters and checkpoints because the frame accuracy metrics in self-sup are not always a good indicator of the final WER performance. To resolve this issue, we propose a joint train for both RNN-T and self-sup loss:
\begin{equation}
\centering
{\cal L} = {\cal L}_{\tt RNN-T} + \lambda {\cal L}_{\tt self-sup}
\label{eq:comb}
\end{equation}
where $\lambda$ is a hyper-parameter. We find $\lambda = 0.5$ works well.

Wav2vec2.0 is a bi-directional method. When we train it with a causal audio encoder, it does not converge. Our workaround is to pre-train with right context and subsequently fine-tune with left context only. Joint Wav2vec2.0 and RNN-T train does not have the issue. The joint train works on a causal audio encoder, because RNN-T loss can guide Wav2vec2.0 objective. As our RNN-T model is online, we use a causal audio encoder for all joint experiments. We add an additional MLP layer between the audio encoder and RNN-T joint layer, and between the audio encoder and self-supervised loss. The representation of the last audio encoder layer is specialized by 2 MLP layers for RNN-T loss and self-supervised loss respectively.

\subsection{Semi-supervised learning}
\label{sec:semisup}
After self-supervised pre-training, we train the ASR model using RNN-T loss with both source domain data (labeled data) and target domain data (unlabeled data) \cite{misra2021comparison}. NST produces pseudo labels for target domain data. 

The teacher model is trained with source domain data (same for the student model). As a result, the pseudo labels generated for the target domain data is error-prone, which is harmful for domain adaptation. When pseudo labels are generated, the teacher model filters out low confidence utterances by Confidence Estimation Module (CEM) \cite{qiu2021learning}. When the teacher model is trained by RNN-T loss, we add CEM whose inputs are the audio encoding and the beam search labels from the RNN-T model. The CEM is trained to minimise the binary cross entropy between the estimated confidence $p$ and the binary target sequence $c$. The target sequence $c$ contains a 1 when the prediction word is correct and 0 otherwise. The average word-level confidence is used to filter utterances. The teacher's multi-task training objective consists of both the RNN-T loss and confidence loss.

\begin{figure}[t]
 \centering
\includegraphics[width=0.38\textwidth]{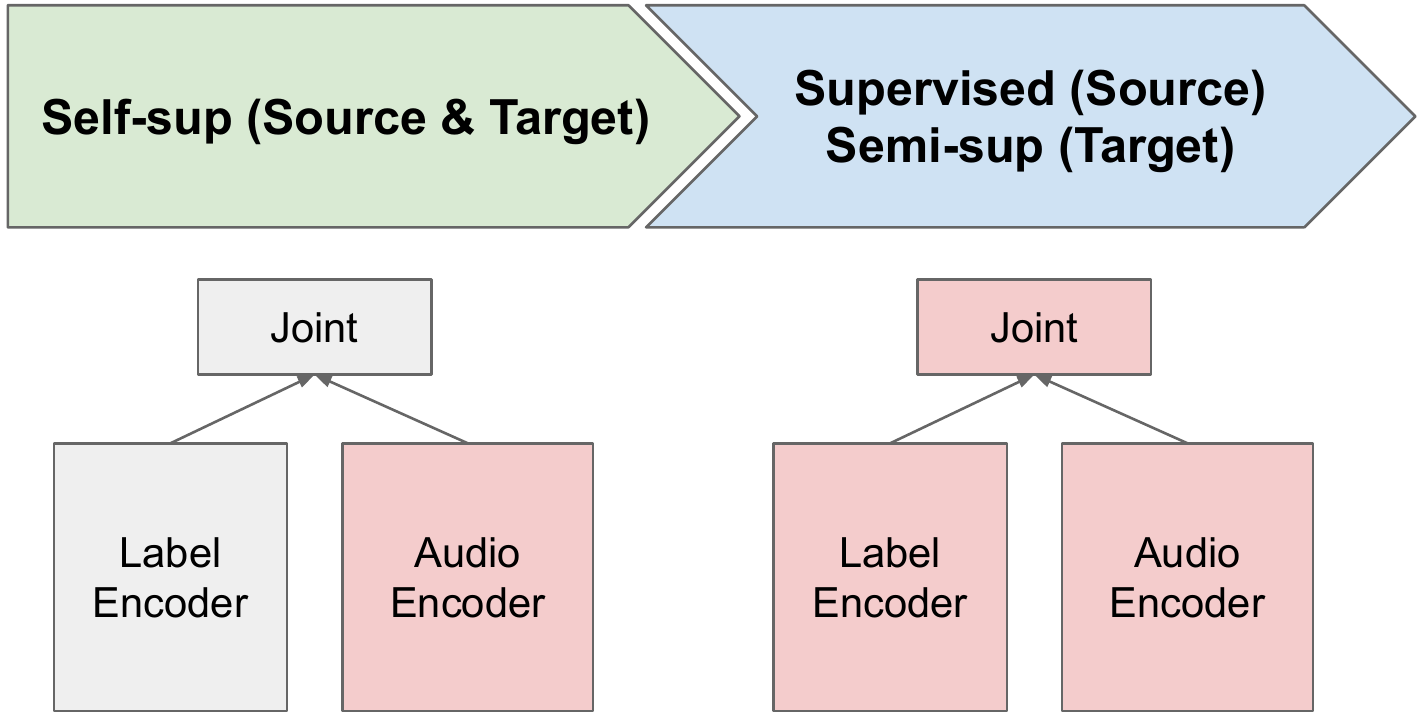}
\caption{Domain adaptation: Self-sup pre-trains the audio encoder. Supervised and semi-sup train all the modules.}
\label{fig:alg}
\vspace{-0.4cm}
\end{figure}

\subsection{Domain adaptation approach}
\label{sec:approach}

Figure \ref{fig:alg} visualizes the proposed domain adaptation method. First of all, self-sup trains the audio encoder with source and target domain data. Then, RNN-T model is trained by RNN-T loss with source and target domain data. NST produces pseudo label for target domain data.

\section{Experimental Setup}
\label{sec:exp}


\subsection{Model Architecture Details}
\label{sec:models}
We use a Conformer RNN-T \cite{Graves2012} model in the experiments. The audio encoder has $17$ Conformer blocks with model dimension $512$. As the model is online ASR, we restrict the  model from using any future information \cite{li2021better}. Each conformer block \cite{DBLP:conf/interspeech/GulatiQCPZYHWZW20,li2021better} uses causal convolution and left-context attention layers. The convolution kernel size is $15$ and the self-attention layer consists of $8$ heads with $65$ left context length. The RNN-T decoder consists of a label encoder with $2$ LSTM layers with $2048$ units projected down to $640$ output units, and a joint network with a single feed-forward layer with $640$ units, as shown in Fig \ref{fig:alg}. The architecture and training procedure is similar to the LibriSpeech SOTA WERs model \cite{zhang2020pushing}. The total number of weights is 137M, while that of the SOTA model is 1B.

The model input is a vector of size $528$, consisting of $4$ contiguous frames of $128$-dimension log-mel features \cite{narayanan2019recognizing} sub-sampled by a factor of $3$ and one-hot domain-id vector of size $16$. The model label uses $4,096$ word pieces \cite{schuster2012japanese} as sub-word label.

\subsection{Data Sets}
\label{sec:data}
We use large multi-domain (\texttt{MD}) data sets \cite{Narayanan2018} in English. \texttt{MD} utterances include multi domain data such as search, farfield, telephony and YouTube. All datasets are anonymized and hand-transcribed except for \texttt{YT} which uses YouTube video transcription with confidence filter \cite{liao2013large}. We divide \texttt{MD} data into source domain data and target domain data. We pre-train the model using source domain data by self-sup and transfer it to target domain data by semi-sup.
As shown in Table \ref{tab:data}, we use Medium-form (\texttt{MF}) utterances as target domain and \texttt{MD-MF} as the source domain. \texttt{MF} is average 10.4 secs length in natural conversation domain. Short-form (\texttt{SF}) is average 4.6 secs length in voice command domain.

For evaluation metrics, we calculate the WER of \texttt{MF} to measure the performance on target domain and the WER of \texttt{SF} for the performance on the source domain. The goal of domain adaptation is to minimize the amount of required transcriptions of \texttt{MF} while maintaining WERs for both \texttt{MF} and \texttt{SF}.
In the final experiments, we use 3\% \texttt{MF} as labeled data, because 800 hours data is manageable amount for hand-transcription. NST produces pseudo labels for rest of 97\% \texttt{MF} data.

\begin{table}[t]
    \centering
    \caption{Overview of training data sets. \texttt{MF} denotes the target domain utterances. \texttt{MD} denotes all domain utterances including \texttt{MF}, \texttt{SF} and \texttt{YT}. \texttt{MD\textsubscript{src}} (\texttt{MD-MF}) denotes the source domain utterances. }
    \label{tab:data}
    \begin{tabular}{ccc}
        \toprule
        \textbf{Data set} & \textbf{Domain} & \textbf{Hours} \\
        \midrule
         Medium-form (\texttt{MF}) & Target & $26$k \\
        \midrule
         Short-form (\texttt{SF})  & Source & $27$k \\
         Youtube (\texttt{YT}) & Source & $226$k \\
        \midrule
         Multi-domain (\texttt{MD}) & Source $\&$ Target & $400$k \\
         \texttt{MD-MF} (\texttt{MD\textsubscript{src}}) & Source & $374$k \\
         3$\%$ \texttt{MF} + \texttt{MD\textsubscript{src}} (\texttt{MD\textsubscript{3p}}) & Source $\&$ Target & $375$k \\
        \bottomrule
    \end{tabular}
    \vspace{-0.4cm}
\end{table}

\section{Experimental results}
\label{sec:results}
In this section, we conduct extensive domain adaptation experiments using self- and semi-supervised learning.

\subsection{Self-supervised Learning}
\label{sec:exp_self}
In Table \ref{tab:exp_self}, we compare the WERs of supervised and self-supervised learning. The first block lists supervised experiments with \texttt{MD}, \texttt{MD\textsubscript{src}}, and \texttt{MD\textsubscript{3p}} data. \texttt{MD} has full \texttt{MF} data. \texttt{MD\textsubscript{src}} doesn't have any \texttt{MF} data. \texttt{MD\textsubscript{3p}} has 3\% \texttt{MF} data. As expected, \texttt{SF} WERs are the same but \texttt{MF} WERs are better when more \texttt{MF} data is used.

The second block lists three different self-sup experiments. One of 3 self-sup methods pre-trains the audio encoder with \texttt{MD}, and RNN-T loss trains the RNN-T model with \texttt{MD\textsubscript{src}}. Wav2vec and APC have better WERs on both \texttt{MF} and \texttt{SF} than Wav2vec2.0, unlike what Wav2vec2.0 paper reported \cite{baevski2020wav2vec}. The downstream ASR model is online RNN-T, which is a causal model. Wav2vec and APC are causal models like GPT-3, but Wav2vec2.0 is full context (non-causal) model like BERT. It shows causal self-sup has better performance for causal downstream task. Even though Wav2vec and APC have the same WERs, we use Wav2vec for rest of experiments. In our experience, APC is more sensitive to checkpoint fluctuations. When we choose a pre-trained checkpoint, Wav2vec works between 50k and 1.2M steps, but APC works only near 100k steps. In addition, APC requires total variation auxiliary loss to stabilize it \cite{huo2021incremental}.

In the third block of Table \ref{tab:exp_self}, Wav2vec \texttt{MD\textsubscript{3p}} enhances WERs for both source domain (\texttt{SF}, 6.0 to 5.8) and target domain (\texttt{MF}, 3.7 to 3.6), compared to Supervised \texttt{MD\textsubscript{3p}}. However, there is huge gap between Supervised \texttt{MD} and Wav2vec \texttt{MD\textsubscript{3p}}. Self-sup enhances overall model generalization, but cannot reduce gap of out-of-domain (OOD) generalization.

\begin{table}[!t]
    \centering
    \caption{Comparisons of Word Error Rate (WER) between supervised baselines, three self-sup algorithms.} 
    \label{tab:exp_self}
    \begin{tabular}{cccc}
        \toprule
        {\textbf{Algorithms}}& {\textbf{Data}} & \multicolumn{2}{c}{\textbf{Word Error Rate ($\%$)}} \\
         & {} & {\textbf{MF (Target)}} & {\textbf{SF (Source)}} \\
        \midrule
        {Supervised} & \texttt{MD} & $\textbf{3.2}$  & {$\textbf{6.0}$} \\
        Supervised & \texttt{MD\textsubscript{src}} & $6.2$  & {$6.0$} \\
        Supervised & \texttt{MD\textsubscript{3p}} & $3.7$  & {$6.0$} \\
        \midrule
        Wav2vec & \texttt{MD\textsubscript{src}} &  ${4.5}$ & {$5.8$} \\
        Wav2vec2.0 & \texttt{MD\textsubscript{src}} & {${4.6}$} & {$5.9$} \\
        APC & \texttt{MD\textsubscript{src}} & ${4.5}$ & {$5.8$} \\
        \midrule
        {Wav2vec} & \texttt{MD} & $\textbf{3.2}$  & {$\textbf{5.8}$} \\
        Wav2vec & \texttt{MD\textsubscript{3p}} &  $\textbf{3.6}$ & {$\textbf{5.8}$} \\
        \bottomrule
    \end{tabular}
    \vspace{-0.4cm}
\end{table}

\subsection{Semi-Supervised Learning}
\label{sec:exp_ssl}

The first block in Table \ref{tab:exp_ssl} shows semi-sup results. Semi-sup \texttt{MD\textsubscript{src}} denotes RNN-T training using \texttt{MD\textsubscript{src}} for labeled data and \texttt{MF} for unlabeled data. Semi-sup \texttt{MD\textsubscript{3p}} denotes RNN-T training using \texttt{MD\textsubscript{3p}} for labeled data and 97\% \texttt{MF} for unlabeled data. NST produces the pseudo labels for unlabeled data. Compared to Supervised \texttt{MD\textsubscript{src}}, Semi-sup \texttt{MD\textsubscript{src}} improves \texttt{MF} (target domain) WER from 6.2 to 3.4. There is only 0.2 \% gap between Supervised \texttt{MD}. Semi-sup can reduce most of gap of OOD generalization, even without target domain data. Semi-sup \texttt{MD\textsubscript{3p}} closes the all the gap, whose \texttt{MF} WER is same to the full data baseline: 3.2\%. All semi-supervised experiments have the same \texttt{SF} WER to the baseline as all of them uses same amount of source domain data.

The teacher model for semi-sup has almost the same architecture as the student model. One difference for the teacher model is that it is non-causal. The self-attention right context is $64$ and convolution layer is not causal. The teacher model uses the same amount of data as the student model. The teacher model is trained with \texttt{MD\textsubscript{src}} for Semi-sup \texttt{MD\textsubscript{src}} and \texttt{MD\textsubscript{3p}} for Semi-sup \texttt{MD\textsubscript{3p}}.

\begin{table}[t]
    \centering
    \caption{WERs for semi-supervised learning.}
    \label{tab:exp_ssl}
    \begin{tabular}{cccc}
        \toprule
        {\textbf{Algorithms}}& {\textbf{Data}} & \multicolumn{2}{c}{\textbf{Word Error Rate ($\%$)}} \\
         & {} & {\textbf{MF (Target)}} & {\textbf{SF (Source)}} \\
        \midrule
        Semi-sup & \texttt{MD\textsubscript{src}} &  ${3.4}$ & {$6.0$} \\
        Semi-sup & \texttt{MD\textsubscript{3p}} &  $\textbf{3.2}$ & {$6.0$} \\
        \midrule
        Self + Semi-sup & \texttt{MD\textsubscript{3p}} &  $\textbf{3.1}$ & {$\textbf{5.7}$} \\
        \bottomrule
    \end{tabular}
    \vspace{-0.4cm}
\end{table}

\vspace{-0.2cm}
\subsection{Self + Semi-Supervised Learning}
\label{sec:exp_self_semi}
The second block in Table \ref{tab:exp_ssl} shows self + semi-sup WERs. In Table \ref{tab:exp_self}, Supervised \texttt{MD} uses all the labels of both source domain (\texttt{MD\textsubscript{src}}) and target domain (\texttt{MF}) data. Wav2vec \texttt{MD\textsubscript{3p}} has better \texttt{SF} (source domain) WER than the supervised baseline because self-sup improves overall generalization as mentioned in Section~\ref{sec:exp_self}. Semi-sup \texttt{MD\textsubscript{3p}} has the same \texttt{MF} (target domain) WER to the baseline because semi-sup resolves OOD generalization as mentioned in Section~\ref{sec:exp_ssl}. Combined both self- and semi-sup are complementary. In Table \ref{tab:exp_ssl}, Self + Semi-sup \texttt{MD\textsubscript{3p}} show even better WERs for both \texttt{MF} (target domain) and \texttt{SF} (source domain). We are actually surprised that that Self + Semi-sup with ${3\%}$ target domain has better WERs than supervised learning with ${100\%}$ target domain. Our domain adaptation method not only reduces all the OOD generalization gap but also improves source domain performance. Semi-sup plays a much more critical role to close the OOD gap.

\subsection{Confidence filter for Semi-Supervised Learning}
\label{sec:exp_conf}

In Table \ref{tab:exp_ssl}, all semi-supervised experiments use a confidence filter \cite{qiu2021learning} that filters out target domain (\texttt{MF}) data when the utterance-level confidence score is less than $0.9$. This results in dropping $24$\% of \texttt{MF} utterances and improves the teacher WER on the target domain by $54$\%. Without confidence filter, both \texttt{MF} and \texttt{SF} WERs are slightly worse.

Using only high-confidence utterances results in higher quality pseudo labeling, which enhances \texttt{MF} (target domain) WER from 3.2 to 3.1. As the teacher is trained with \texttt{MD\textsubscript{3p}}, it tends to be more confident to source-domain-like \texttt{SF} data. This results in additional \texttt{SF} (source domain) WER improvement from 5.8 to 5.7. It indicates that confidence score threshold should be tuned per fraction of target label data.

\begin{table}[ht]
    \centering
    \caption{WERs of combined self- and semi-sup with/without confidence filter on \texttt{MD\textsubscript{3p}}.}
    \label{tab:conf_filter}
    \begin{tabular}{cccc}
        \toprule
        {\textbf{Algorithms}}& \multicolumn{2}{c}{\textbf{Word Error Rate ($\%$)}} \\
         &  {\textbf{MF (Target)}} & {\textbf{SF (Source)}} \\
        \midrule
        Self + Semi-sup w/o filter &  ${3.2}$ & {$5.8$} \\
        Self + Semi-sup w filter &  ${3.1}$ & {$5.7$} \\
        \bottomrule
    \end{tabular}
    \vspace{-0.6cm}
\end{table}

\subsection{Joint RNNT and Self-Sup train}
\label{sec:exp_comb}

In Fig \ref{fig:combined}, Joint W2V2 (Green line) reaches WER 4.6 at 200k step, while W2V2 requires 900k steps for both pre-training and supervised learning. Total training steps is reduced significantly. In addition, Joint W2V2 reaches WER 4.3, which W2V2 never reach. It's because Joint W2V2 utilizes Wav2vec2.0 loss on the causal encoder, so there is no transition gap from non-causal to causal unlike the Wav2vec2.0 experiment. It brings out the full potential of Wav2vec2.0 objective. However, Joint W2V (Purple line) requires the same steps to reach the converging point to W2V (Blue line), as both utilize a causal audio encoder.

In Table \ref{tab:combined}, Joint W2V + NST and Joint W2V2 + NST has additional semi-sup stage. They have the same \texttt{MF} (target domain) WER to Self + Semi-sup \texttt{MD\textsubscript{3p}} in Table \ref{tab:exp_ssl}. However, their \texttt{SF} (source domain) WER is worse than Self + Semi-sup \texttt{MD\textsubscript{3p}} (5.7\%). We leave the weaker in-domain generalization power of Joint train for future study.

\begin{figure}[ht]
 \centering
\includegraphics[width=0.48\textwidth]{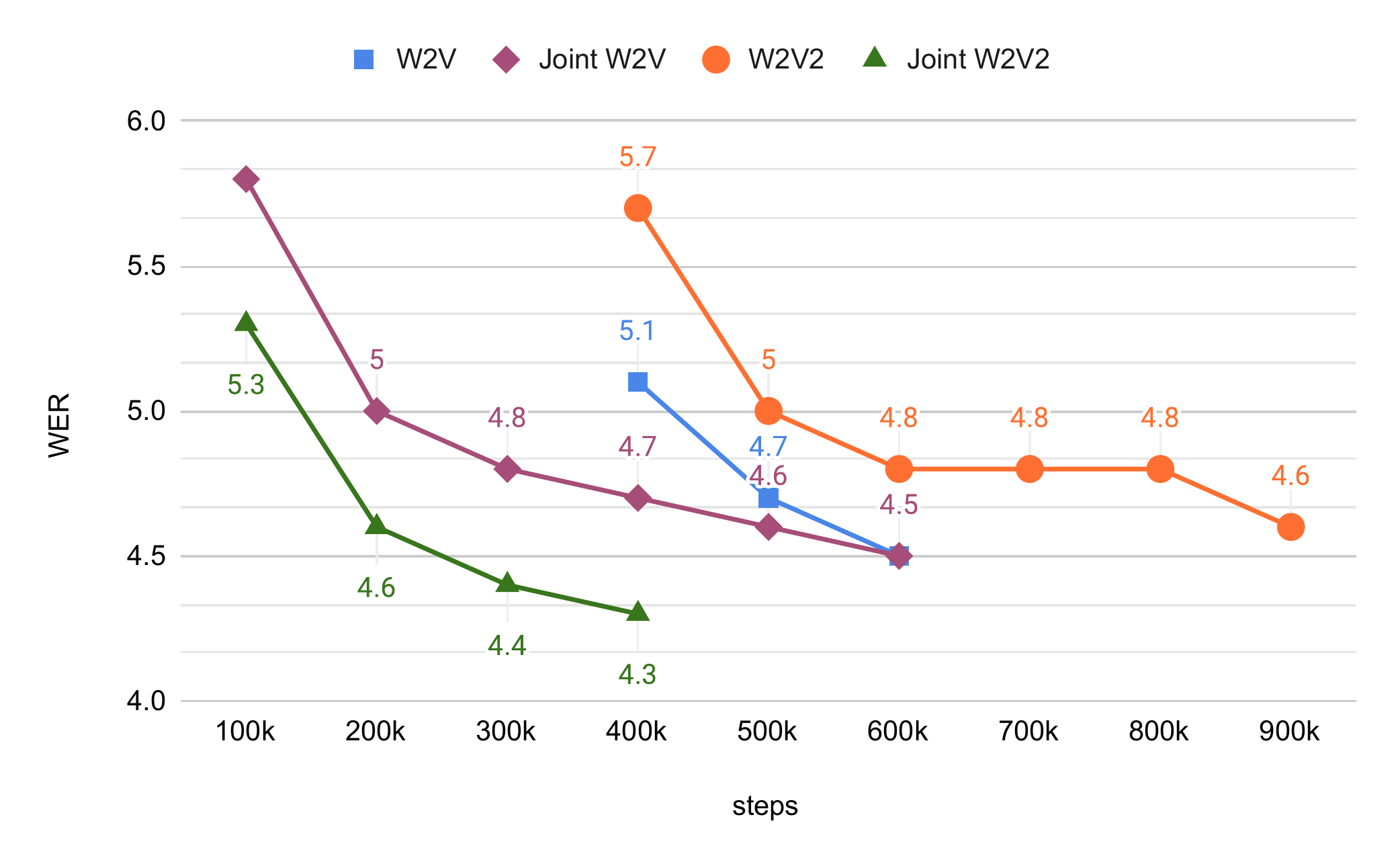}
\caption{Comparison of WER convergence with and without Joint train for wav2vec and wav2vec2.0. }
\label{fig:combined}
\vspace{-0.4cm}
\end{figure}

\begin{table}[ht]
    \centering
    \caption{WERs of Joint train with \texttt{MD\textsubscript{3p}}.}
    \label{tab:combined}
    \begin{tabular}{cccc}
        \toprule
        {\textbf{Algorithms}}& \multicolumn{2}{c}{\textbf{Word Error Rate ($\%$)}} \\
         &  {\textbf{MF (Target)}} & {\textbf{SF (Source)}} \\
        \midrule
        Joint W2V &  ${4.5}$ & {$5.8$} \\
        Joint W2V2 &  ${4.3}$ & {$5.8$} \\
        Joint W2V + NST &  $\textbf{3.1}$ & {$5.9$} \\
        Joint W2V2 + NST &  $\textbf{3.1}$ & {$5.8$} \\
        \bottomrule
    \end{tabular}
    \vspace{-0.6cm}
\end{table}

\section{Conclusion}
\label{sec:conclusion}
In this paper, we investigated a domain adaptation method by combining self- and semi-supervised learning. Self-sup improves overall generalization and semi-sup closes out-of-domain generalization gap. Both methods complement each other. Extensive experimental results demonstrate that the proposed methods using only 3\% of the labeled target domain data obtain better WERs on both the target and source domains than the baseline.

\section{Acknowledgements}

We thank Hasim Sak, Anshuman Tripathi, Jaeyoung Kim, Qian Zhang, Yonghui Wu, Ruoming Pang, Yu Zhang, Arun Narayanan, Tony Bruguier, Lillian Zhou, Petr Zadrazil, Zhehuai Chen, Andrew Rosenberg for helpful discussions.


\bibliographystyle{IEEEbib}
\bibliography{refs}

\newpage 
\appendix

\section{Appendix}

\subsection{Self-sup with X$\%$ \textbf{MF} data}
Self-sup enhances WERs when labeled data is less than ${10\%}$ (low resource regime, less than 2.6k hours). However, self-sup has diminishing returns in high resource regime.

\begin{figure}[ht]
 \centering
\includegraphics[width=0.48\textwidth]{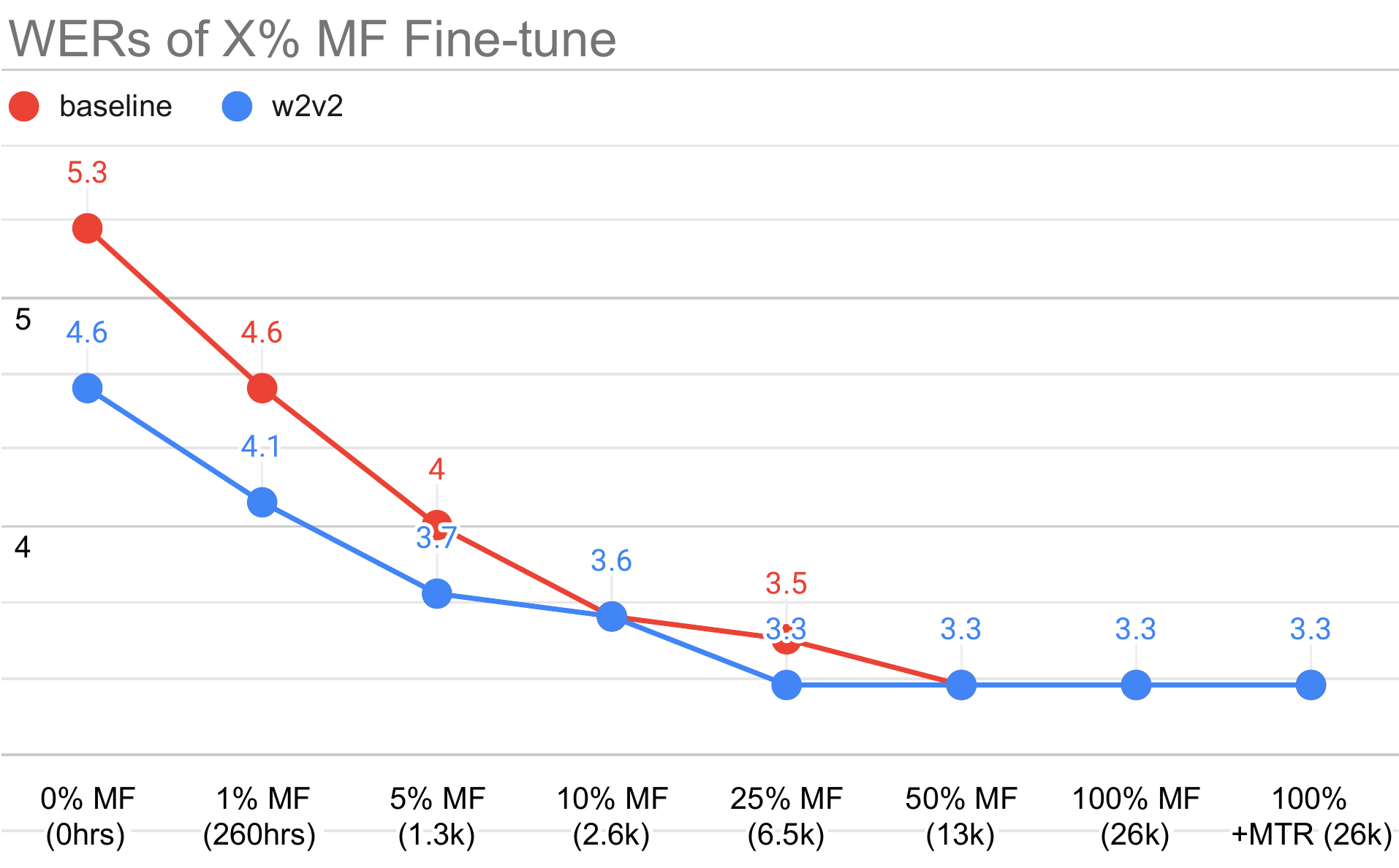}
\caption{Comparison of WER on X$\%$ MF data between supervised baseline and Wav2vec. }
\label{fig:combined}
\vspace{-0.4cm}
\end{figure}

\subsection{Semi-sup with full \textbf{MD} data}

In semi-sup training, ${3\%}$ label (i.e. ${97\%}$ pseudo label) and ${0\%}$ label don't make any difference. Pseudo label is as good as human transcription. It means we need only ${3\%}$ label with semi-sup.

\begin{table}[ht]
    \centering
    \caption{WERs for semi-supervised learning.}
    \label{tab:exp_ssl_full}
    \begin{tabular}{cccc}
        \toprule
        {\textbf{Algorithms}}& {\textbf{Data}} & \multicolumn{2}{c}{\textbf{Word Error Rate ($\%$)}} \\
         & {} & {\textbf{MF (Target)}} & {\textbf{SF (Source)}} \\
        \midrule
        Semi-sup & \texttt{MD\textsubscript{src}} &  ${3.4}$ & {$6.0$} \\
        Semi-sup & \texttt{MD\textsubscript{3p}} &  $\textbf{3.2}$ & {$6.0$} \\
        Semi-sup & \texttt{MD} &  $\textbf{3.2}$ & {$6.0$} \\
        \midrule
        Self + Semi-sup & \texttt{MD\textsubscript{3p}} &  $\textbf{3.1}$ & {$\textbf{5.7}$} \\
        Self + Semi-sup & \texttt{MD} &  $\textbf{3.1}$ & {$\textbf{5.7}$} \\
        \bottomrule
    \end{tabular}
    \vspace{-0.4cm}
\end{table}

\end{document}